\documentclass[aps,11pt,nofootinbib,groupedaddress,preprintnumbers,superscriptaddress]{revtex4-1}

\usepackage{amssymb, amsfonts, amsmath, bm}
\usepackage%[dvipdfmx]
{graphicx}
\usepackage{color}
\usepackage%[dvipdfmx]
{hyperref}
\hypersetup{% 
 setpagesize=false,
 bookmarksnumbered=true,%
 bookmarksopen=true,%
 colorlinks=true,%
 linkcolor=blue,%
 citecolor=red}
\begin{document}
\title{
Stable circular orbits in higher-dimensional multi-black hole spacetimes
}
\author{{\large Takahisa Igata}}
\email{igata@post.kek.jp}
\affiliation{KEK Theory Center, 
Institute of Particle and Nuclear Studies, 
High Energy Accelerator Research Organization,Tsukuba 305-0801, Japan}
\author{{\large Shinya Tomizawa}}
\email{tomizawa@toyota-ti.ac.jp}
\affiliation{
Mathematical Physics Laboratory, 
Toyota Technological Institute, Nagoya 468-8511, Japan}
\date{\today}
\preprint{KEK-TH-2248}
\preprint{KEK-Cosmo-262}
\preprint{TTI-MATHPHYS-1}

\begin{abstract}
We consider the dynamics of particles, particularly focusing on circular orbits 
in the higher-dimensional Majumdar-Papapetrou (MP) spacetimes 
with two equal mass black holes. 
It is widely known that in the 5D Schwarzschild-Tangherlini and Myers-Perry backgrounds, 
there are no stable circular orbits. In contrast, we show that in the 5D MP background, 
stable circular orbits can always exist 
when the separation of two black holes is large enough. 
More precisely, for a large separation, stable circular orbits exist 
from the vicinity of horizons to infinity; for a medium one, 
they appear only in a certain finite region bounded 
by the innermost stable circular orbit 
and the outermost stable circular orbit outside the horizons; 
for a small one, they do not appear at all. 
Moreover, we show that in MP spacetimes in more than 5D, they do not exist for any separations.
\end{abstract}

\maketitle

%%%%%%%%%%
\section{Introduction}
\label{sec:1}
%%%%%%%%%%

The dynamics of test free particles in curved spacetimes, i.e., 
the geodesic structure, include important information about the 
gravitational field and the geometry. 
In stationary spacetimes, there can be stationary orbits of particles, 
which are geodesics along timelike Killing fields. 
Furthermore, if it is also axisymmetric, the stationary orbits can be circular orbits.
Such fundamental orbits associated with spacetime symmetries are useful 
to understand various observable phenomena 
(e.g., the stellar motion and black hole shadow) around the black hole.

In the Schwarzschild black hole spacetime, 
there exist both stable and unstable circular orbits of particles. 
Let $r$ be the circumference radius, and let $M$ be the black hole mass. 
We know that the stable circular orbits exist in the range $r\geq 6M$, and 
the unstable circular orbits in the range $3M<r<6M$, where 
we have used geometrized units. 
There exist the innermost stable circular orbit (ISCO) at their boundary $r=6M$ and 
the unstable photon circular orbit at the last circular orbit $r=3M$. 
These are fundamentals of physical phenomena in the vicinity of a black hole.
In the Kerr black hole spacetime, 
both stable and unstable circular orbits also appear~\cite{Wilkins:1972rs}.

In the last two decades, higher-dimensional black holes have also been 
actively studied~\cite{Emparan:2008eg}, 
so that some of them are parametrized by the spacetime dimension $d$. 
The parametrization allows us to distinguish 
between $d$-dependent/independent properties 
and also tells us some special properties in a specific $d$. 
Such dimensionality often appears in the analysis of gravitational properties 
through geodesic structure, which is the first step in the study of black holes.
Unlike the 4D case, 
in a higher-dimensional static and spherically symmetric vacuum black hole, 
there is no stable circular orbit 
because no stable balance is formed 
between the gravitational and centrifugal forces~\cite{Tangherlini:1963bw}, 
which is a generic feature of higher-dimensional black holes 
with a spherically symmetric horizon~\cite{Hackmann:2008tu}. 
This property carries over to 
circular orbits in the 5D Myers-Perry 
black holes~\cite{Frolov:2003en, Kagramanova:2012hw, Diemer:2014lba} 
and equatorial circular orbits in the singly rotating Myers-Perry black holes 
in arbitrary dimensions~\cite{Cardoso:2008bp}. 
Though it was pointed out that 
stable stationary/bound orbits can exist in the Myers-Perry black holes~\cite{Igata:2014xca} 
at least when there is no upper limit on the black hole spin parameters 
(so-called the ultraspinning limit~\cite{Emparan:2003sy}),%
\footnote{In higher-dimensional AdS black holes, 
stable stationary orbits can appear because of the asymptotic structure~\cite{Delsate:2015ina, Grunau:2017uzf}.} 
they tend not to appear for the higher-dimensional Myers-Perry family in general 
due to the dimensional dependence of the law of gravity.

However, there seems to be some exceptions to the nature that stable circular/bound orbits 
are less likely to appear under the diversity of higher-dimensional gravity.
As one of the rich properties of higher-dimensional spacetimes, 
there is the topological variety of spatial cross sections of horizons. 
In 5D asymptotically flat, stationary, and biaxisymmetric spacetimes, 
the allowed horizon topology is not only the sphere $S^3$ 
but also the ring $S^1\times S^2$ and the lens 
$L(p, q)$~\cite{Hollands:2007aj,Hollands:2010qy,Cai:2001su,Galloway:2005mf}. 
We are gradually learning that the nontrivial horizon topologies of black objects 
can give rise to a mechanism for the appearance of stable stationary orbits 
that is different from the case of spherical black holes.
In 5D black ring spacetimes~\cite{Emparan:2001wn}, 
stable stationary orbits are absent in the fat regime, as is the spherical cases, 
but appears in the thin regime~\cite{Hoskisson:2007zk, Igata:2010ye, Grunau:2012ai, Igata:2013be}. 
The existence of a nut%
\footnote{This terminology is often used to denote an isolated fixed point of 
one-parameter $U(1)$ isometry~\cite{Gibbons:1979xm}.}
outside the horizon plays an essential role in the 
existence of stable stationary orbits. 
Recently, even for the black rings in more than 5D, 
the existence of stable stationary orbits and their dimensionality have been revealed 
using the blackfold approach~\cite{Igata:2020vdb, Igata:2020dow}. 
In 5D black lens spacetimes~\cite{Kunduri:2014kja, Tomizawa:2016kjh}, 
stable circular orbits can also exist~\cite{Tomizawa:2019egx}. 
Even in this phenomenon, it is also essential that the centers 
(i.e., the nuts) are located outside the horizon.

Concerning a higher-dimensional black hole with disconnected components 
of the horizon cross section, 
we encounter a nontrivial question of whether stable stationary/bound orbits exist and, 
if so, how they are distributed. 
This paper aims to clarify how the many-body nature of black holes 
in higher-dimensional spacetimes affects the existence of test particles' stationary orbits.
To explore this, we adopt the two-body black hole configuration of 
the Majumdmar-Papapetrou~(MP) 
geometry~\cite{Majumdar:1947eu,Papaetrou:1947ib, Myers:1986rx}, 
which is kept static by balancing the gravitational and Coulomb force 
between the two with electric charges of the same sign. 
Since this family has singularities on the horizon but not outside it, 
we can examine the existence of the timelike/null geodesic Killing orbits 
and their stability throughout the domain of outer communication. 
The geodesic structure of the 4D MP dihole spacetime has been analyzed in detail 
in terms of circular/bound orbits and 
their stability~\cite{Chandrasekhar:1989vk, Contopoulos:1990, 
Wunsch:2013st, Dolan:2016bxj, Ono:2016lql, Assumpcao:2018bka, 
Nakashi:2019mvs, Nakashi:2019tbz}, 
chaos~\cite{Contopoulos:1991, Shipley:2016omi}, 
and shadows~\cite{Nitta:2011in, Patil:2016oav}.
The particle dynamics in higher-dimensional MP spacetimes were 
investigated in Ref.~\cite{Hanan:2006uf}. 
Since the center is located outside the horizon 
in a higher-dimensional two-body black hole spacetime, 
as well as the black rings and black lenses, 
we can expect the appearance of stable stationary orbits 
in the higher-dimensional MP dihole spacetimes.

This paper is organized as follows. 
In Sec.~\ref{sec:2}, we introduce the MP dihole spacetime 
in $d$ dimensions and formulate particle dynamics on the spacetime. 
Focusing specifically on stationary orbits, 
we identify the conditions for their existence 
and clarify criteria to determine whether they are stable.
In Sec.~\ref{sec:3}, for $d=5$, 
we clarify the dependence of the sequences of stationary orbits 
on the dihole separation parameter. 
In addition, based on these results, 
we give some critical values for the separation parameter.
We discuss these properties for $d\geq 6$ as well.
Section~\ref{sec:4} is devoted to a summary and discussions. 
Throughout this paper we use units in which $G=1$ and $c=1$.

%%%%%%%%%%
\section{Formulation}
\label{sec:2}
%%%%%%%%%%
We focus on the MP geometries in $d$ dimensions~($d\geq 4$). The metric and the gauge field are given by%
\footnote{This is a solution in the $d$-dimensional Einstein-Maxwell theory, whose action is given by
\begin{align}
S=\frac{1}{16\pi G} \int \mathrm{d}^d x \sqrt{-g} (R-F_{\mu\nu}F^{\mu\nu}),
\end{align}
where $R$ is the Ricci tensor, and $F_{\mu\nu}$
is the field strength of the gauge field, and we have restored the $d$-dimensional Newton constant $G$. 
%$G_{\mu\nu}=8\pi T_{\mu\nu}$, 
}
\begin{align}
g_{\mu\nu}\:\!\mathrm{d}x^\mu \:\!\mathrm{d}x^\nu
&=-U^{-2}\:\!\mathrm{d}t^2+U^{2/(d-3)}
\mathrm{d}\bm{r}\cdot \mathrm{d} \bm{r},
\\
A_\mu \:\!\mathrm{d}x^\mu&=\sqrt{\frac{d-2}{2(d-3)}} U^{-1}\:\!\mathrm{d}t,
\end{align}
where $\mu$, $\nu$ are spacetime indices, 
and $\mathrm{d}\bm{r}\cdot \mathrm{d} \bm{r}$ is the $(d-1)$-dimensional flat metric, 
and $U$ is a harmonic function on $\mathbb{R}^{d-1}$~\cite{Majumdar:1947eu, 
Papaetrou:1947ib, Myers:1986rx}. 
When $U$ has two point sources, 
the geometry represents a two-centered black hole spacetime. 
Using a $d$-dimensional cylindrical coordinate system 
$(z, \rho, \phi_1,\ldots, \phi_{d-3})$ on $\mathbb{R}^{d-1}$, 
where $z$ is a cylindrical and Cartesian coordinate, $\rho$ is a radial coordinate 
from the $z$ axis, and $\phi_a$ ($a=1, \ldots, d-3$) are polar coordinates orthogonal to 
the $\rho$-$z$ plane, 
the metric of the MP dihole spacetime is given by
\begin{align}
\label{eq:met}
g_{\mu\nu}\:\!\mathrm{d}x^\mu \:\!\mathrm{d}x^\nu
&=-U^{-2}\:\!\mathrm{d}t^2+U^{2/(d-3)}\left(
\mathrm{d}z^2+\mathrm{d}\rho^2+\rho^2\:\!\mathrm{d}\Omega^2_{d-3}
\right),
\\
U&=1+\frac{M_+}{r_+^{d-3}}+\frac{M_-}{r_-^{d-3}},
\\
\label{eq:rpm}
r_{\pm}&=\sqrt{(z\pm a)^2+\rho^2},
\end{align}
where $M_\pm$ are masses of two extremal black holes placed at $z=\mp a$ on the $z$ axis, 
and $\mathrm{d}\Omega_{d-3}^2$ is the metric on the unit $S^{d-3}$. 
We assume that the two black holes have equal mass, $M_+=M_-=M$, in what follows.

We focus on particle dynamics in the dihole spacetime. 
Let $p_{\mu}$ be canonical momenta conjugate to coordinates, $x^\mu$. 
The Hamiltonian of a freely falling particle with unit/zero mass is given by 
\begin{align}
H=\frac{1}{2} g^{\mu\nu}p_\mu p_\nu=\frac{1}{2}\left[\:\!
-U^2 p_t^2+U^{-2/(d-3)} \left(
p_z^2+p_\rho^2+\frac{1}{\rho^2}\gamma^{ab}p_ap_b
\right)
\:\!\right],
\end{align}
where $g^{\mu\nu}$ is the inverse metric of $g_{\mu\nu}$, and 
$\gamma^{ab}$ is the inverse of the metric on $S^{d-3}$. 
The momentum $p_t=-E$ is a conserved energy because 
$H$ is independent of time $t$. 
The quadratic quantity $\gamma^{ab}p_ap_b=L^2$ is also 
a constant of motion associated with spherical symmetry on $S^{d-3}$.

We consider stationary orbits on which a particle takes constant $z$ and $ \rho$. 
The on-shell condition of geodesic motion $g^{\mu\nu}p_\mu p_\nu+\kappa=0$, 
where $\kappa$ is squared particle mass, yields
\begin{align}
&U^{2(4-d)/(d-3)}(\dot{z}^2+\dot{\rho}^2)
+
V=E^2,
\\
&V(\rho, z; L^2)=
\frac{L^2}{\rho^2 U^{2(d-2)/(d-3)}}+\frac{\kappa}{U^2},
\end{align}
where the dots denote the derivatives with respect to an affine parameter along the geodesic.
We call $V$ the effective potential. 
Let us focus on particles with $\kappa=1$ 
staying in stationary orbits. 
The conditions of the stationary orbits for $V$ 
and $V_i:=\partial_iV$ ($i=z, \rho$) are written as 
\begin{align}
\label{eq:Vz}
V_z&=-\frac{2\:\!U_z}{U^3}\left(
\frac{d-2}{d-3} \frac{L^2}{\rho^2\:\! U^{2/(d-3)}}+\kappa
\right)
=0,
\\
\label{eq:Vrho}
V_\rho&=-\frac{2L^2}{\rho^3 U^{2(d-2)/(d-3)}}-\frac{2\:\!U_\rho}{U^3} \left(
\frac{d-2}{d-3} \frac{L^2}{\rho^2 U^{2/(d-3)}}+\kappa
\right)=0,
\\
\label{eq:V=E2}
V&=E^2,
\end{align}
respectively, where $U_i:=\partial_iU$ ($i=z, \rho$) take the forms
\begin{align}
U_z&=-(d-3)M \left(
\frac{z+a}{r_+^{d-1}}+\frac{z-a}{r_-^{d-1}}
\right),
\\
U_\rho&=-(d-3) M\rho \left(\frac{1}{r_+^{d-1}}+\frac{1}{r_-^{d-1}}\right),
\end{align}
respectively. 
Solving the condition~\eqref{eq:Vrho} for $L^2$, we have
\begin{align}
\label{eq:L=L0}
L^2=L_0^2:=-\frac{(d-3) \:\!\rho^3 U_\rho U^{2/(d-3)}}{(d-3)U+(d-2)\rho \:\!U_\rho}.
\end{align}
Note that $L_0^2$ must not be negative, 
which is a necessary condition for the existence of a stationary orbit.
From the condition~\eqref{eq:V=E2} together with $L_0^2$, we obtain
\begin{align}
E^2=E^2_0:=V(\rho, z; L_0^2)=\frac{(d-3) U+\rho\:\! U_\rho}{U^2 \left[\:\!
(d-3)U+(d-2) \rho \:\!U_\rho
\:\!\right]},
\end{align}
which also must not be negative. 
Note that if $L_0^2\geq 0$, we have $E_0^2>0$. 
The condition~\eqref{eq:Vz} implies $U_z=0$, 
which defines curves on the $\rho$-$z$ plane. 
These curves are distributed in the range $|z|<a$ 
and always include $z=0$. 
Now we define a family of the curves on the $\rho$-$z$ plane satisfying $L_0^2\geq0$,
\begin{align}
\gamma_0:=\{\:\!(\rho, z)\:\!|\:\! U_z=0, L_0^2\geq 0\:\!\},
\end{align}
which provides the sequence of stationary orbits. 
If evaluated at points on $\gamma_0$, the quantities $L_0$ and $E_0$ give 
the angular momentum and en
ergy of a particle in stationary orbits, respectively. 
Note that all of the stationary orbits we are considering here are circular orbits. 
This is because, due to the spherical symmetry on $S^{d-3}$, 
particles in stationary orbits always move geodesically 
along a certain great circle on the sphere.
Therefore, we will call the stationary orbit a circular orbit.

Now we look for a subset of $\gamma_0$ in which the stationary orbits are stable. 
Let $(V_{ij})$ be the Hessian matrix of $V$, where 
$V_{ij}:=\partial_j \partial_i V$. 
We define $h$ and $k$ as the determinant and the trace of $(V_{ij})$, i.e., 
$h(\rho, z; L^2):=\mathrm{det}(V_{ij})$ and 
$k(\rho, z; L^2):=\mathrm{tr} (V_{ij})$, respectively. 
In terms of $h$ and $k$, we define the region $D$ in which the circular orbits are stable as
\begin{align}
D:=\{\:\!
(\rho, z)\:\!|\:\! h_0>0, k_0>0, L_0^2 \geq 0
\:\!\},
\end{align}
where $h_0$ and $k_0$ are defined by
\begin{align}
h_0&:=\left.h(\rho, z; L_0^2)\right|_{U_z=0},
\\
k_0&:=\left.k(\rho, z; L_0^2)\right|_{U_z=0},
\end{align}
respectively. The restriction denoted by $U_z=0$ means that 
we have directly dropped the terms including $U_z$. 
Thus, we can visualize the sequence of stable circular orbits by 
$\gamma_0$ and $D$ in the $\rho$-$z$ plane. 
Thus, we can visualize the sequence of stable circular orbits by 
the overlap of $\gamma_0$ and $D$ in the $\rho$-$z$ plane. 
Note that $D$ only serves to find the subset of $\gamma_0$.

Here, we summarize the quantities
$E_0$, $L_0$, $h_0$, and $k_0$ evaluated on $z=0$. 
We introduce the $(d-2)$-dimensional radial coordinate defined by
$R:=\sqrt{\rho^2+a^2}$ for simplification of both 
calculations and expressions, 
where note that
$R\geq a$. Let us use units in which $M=1$ in what follows. 
The energy and angular momentum of a particle in a circular orbit on $z=0$ are given by
\begin{align}
\label{eq:E0}
E_0^2(\rho, 0)&=\frac{R^{2(d-3)}(R^{d-1}+2a^2)}{
(R^{d-3}+2)^2 f},
\\
\label{eq:L0}
L_0^2(\rho, 0)&=\frac{2(d-3)(R^2-a^2)^2(R^{d-3}+2 )^{2/(d-3)}}{R^2f},
\end{align}
respectively, where 
\begin{align}
f(R):=
R^{d-1}-2(d-3) R^2+2(d-2)a^2.
\end{align}
Note that $f(R)$ must be always positive on $\gamma_0$. 
The derivatives of $E_0(\rho, 0)$ and $L_0(\rho, 0)$
with respect to $R$ are given by
\begin{align}
\label{eq:dE0}
\frac{\mathrm{d}E_0(\rho, 0)}{\mathrm{d}R}
&=\frac{(d-3)\:\! g}{R(R^{d-3}+2)(R^{d-1}+2\:\!a^2) f},
\\
\label{eq:dL0}
\frac{\mathrm{d}L_0(\rho, 0)}{\mathrm{d}R}
&=\frac{g}{2 R(R^{d-3}+2)(R^2-a^2) f},
\end{align}
respectively, where 
\begin{align}
g(R):=8(d-2)a^4+\left[\:\!
2(3d-1)R^{d-1}
+(d-1)R^{2(d-2)}
-8(d-4)R^2
\:\!\right]a^2
\cr
-6(d-3) R^{d+1}
-(d-5) R^{2(d-1)}.
\end{align}
The signs of these derivatives on $\gamma_0$ are determined by that of $g(R)$. 
The quantities $h_0$ and $k_0$ 
evaluated on $z=0$ are given by
\begin{align}
\label{eq:h0}
h_0(\rho, 0)
&=\frac{16(d-3)^2R^{2(2d-9)}\left[\:\!
R^2-(d-1)a^2\:\!\right] g}{(R^{d-3}+2)^6
f^2},
\\
\label{eq:k0}
k_0(\rho, 0)
&=\frac{4(d-3)R^{2(d-5)}\left[\:\!
g+R^2(R^{d-3}+2)^2\left[\:\!
R^2-(d-1)a^2\:\!\right]
\:\!\right]}{
(R^{d-3}+2)^4
f},
\end{align}
respectively.

Now let us discuss some $d$-independent properties. 
One property common to these systems is that the center of the system 
(i.e., the center of the two black holes) is located outside the horizon. 
This fact leads to a common property in the structure of $V$. 
On $z=0$, the expansion of $V$ around $\rho=0$ is written as
\begin{align}
V(\rho, 0)=\frac{a^{2(d-2)} L^2}{(a^{d-3}+2)^{2(d-2)/(d-3)} \rho^2}+O(\rho^0).
\end{align}
Note that the power of $\rho$ in the leading term does not depend on $d$. 
If $L\neq0$, then $V(\rho, 0)$ diverges in the limit $\rho\to 0$, 
which shows the appearance of the centrifugal barrier near the center. 
Since gravitational force acts attractively, there always exists a stable balance 
between the gravitational force and the centrifugal force 
in the $\rho$ direction near the center.
In the $z$ direction, 
$V$ makes a local maximum in the range $a\leq R<a \sqrt{d-1}$ because 
$V_z(\rho,0)=0$ from the reflection symmetry, 
and $V_{zz}(\rho, 0)>0$, where 
\begin{align}
V_{zz}(\rho, 0)=\frac{4\left[\:\!
R^2-(d-1)a^2
\:\!\right]}{(R^2-a^2)(R^{d-3}+2)^3}\left[\:\!
\frac{(d-2) L^2 R^{2(d-4)}}{(R^{d-3}+2)^{2/(d-3)}}+
(d-3) R^{2(d-5)}(R^2-a^2)
\:\!\right].
\end{align}
Hence, $V$ always makes a saddle point near the center, so that 
no stable circular orbits appear there.

%%%%%%%%%%
\section{Stable/unstable circular orbits}
\label{sec:3}
%%%%%%%%%%

%%%%%
\subsection{$d=5$}
%%%%%
We consider how the sequence of circular orbits varies as the dihole separation 
gradually decreases from a sufficiently large value in the 5D MP spacetime. 
We first check the explicit forms of quantities evaluated on the symmetric plane $z=0$. 
From Eqs.~\eqref{eq:E0} and \eqref{eq:L0}, 
$E_0^2$ and $L_0^2$ in $d=5$ are given by
\begin{align}
\label{eq:E05}
E_0^2(\rho, 0)&=\frac{R^4(R^4+2a^2)}{(R^2+2)^2f},
\\
\label{eq:L05}
L_0^2(\rho, 0)&=\frac{4(R^2-a^2)^2(R^2+2)}{R^2f},
\end{align}
respectively, where 
\begin{align}
\label{eq:f5}
f(R)=R^4-4R^2+6a^2.
\end{align}
From Eqs.~\eqref{eq:dE0} and \eqref{eq:dL0}, 
the derivatives of $E_0(\rho, 0)$ and $L_0(\rho, 0)$ take the forms 
\begin{align}
\frac{\mathrm{d}E_0(\rho, 0)}{\mathrm{d}R}
&=\frac{2 g}{R(R^2+2)(R^4+2\:\!a^2) f},
\\
\frac{\mathrm{d}L_0(\rho, 0)}{\mathrm{d}R}
&=\frac{g}{2 R(R^2+2)(R^2-a^2) f},
\end{align}
respectively, where 
\begin{align}
g(R)=4\left[\:\!
(a^2-3)R^6+R^2(R^2-4\:\!a^2)+(7a^2-1)R^4+2\:\!a^2(R^2+3\:\!a^2)
\:\!\right]. 
\end{align}
From Eqs.~\eqref{eq:h0} and \eqref{eq:k0}, the quantities $h_0$ and $k_0$ take the forms 
\begin{align}
h_0(\rho, 0)&=\frac{64 R^2(
R^2-4a^2)g
}{(R^2+2)^6f^2},
\\
k_0(\rho, 0)&=\frac{8\left[\:\!
g+R^2(R^2+2)^2(R^2-4a^2)
\:\!\right]}{(R^2+2)^4 f},
\end{align}
respectively. 
Besides $z=0$, there is the following branch of $U_z=0$: 
\begin{align}
\label{eq:z0}
z=z_0(R):=\pm \sqrt{R(2\:\!a-R)},
\end{align}
where $a<R\leq 2a$ (i.e., $0<\rho\leq \sqrt{3} a$). 
The particle energy and angular momentum in a circular orbit on $z=z_0$ are given by
\begin{align}
\label{eq:E0z0}
E_0^2(\rho, z_0)
&=\frac{4\:\!a^2(R-a)^3(4\:\!aR+1)}{
\left[\:\!2a(R-a)+1\:\!\right]^2
F},
\\
\label{eq:L0z0}
L_0^2(\rho, z_0)&=\frac{\left[\:\!
2a(R-a)+1
\:\!\right](R+a)^2}{aF},
\end{align}
respectively, where
\begin{align}
F(R):=4\:\!a R^2-(4\:\!a^2+1) R-3\:\!a.
\end{align}
Note that $F(R)$ must be always positive on $\gamma_0$. 
The derivatives of $E_0(\rho, z_0)$ and $L_0(\rho, z_0)$ with respect to $R$ are given by
\begin{align}
\label{eq:dE0c/dR}
\frac{\mathrm{d}E_0(\rho, z_0)}{\mathrm{d} R}
&=
\frac{G}{(R-a) \left[\:\!
2\:\!a (R-a)+1
\:\!\right](4\:\!a R+1 ) F},
\\
\label{eq:dL0c/dR}
\frac{\mathrm{d}L_0(\rho, z_0)}{\mathrm{d} R}
&=\frac{G}{2(R+a)\left[\:\!
2\:\!a(R-a)+1
\:\!\right]F},
\end{align}
respectively, where
\begin{align}
G(R):=8\:\!a^2(R-a)^3-4a\:\!R^2-(28\:\!a^2+1) R+a(8\:\!a^2-5). 
\end{align}
The signs of these quantities on $\gamma_0$ are determined by that of $G(R)$ . 
The quantities $h_0$ and $k_0$ take the forms
\begin{align}
h_0(\rho, z_0)&=\frac{128\:\!a^2(R-a)^3(2\:\!a-R)G}{RF^2\left[\:\!
2\:\!a(R-a)+1
\:\!\right]^6},
\\
k_0(\rho, z_0)
&=\frac{8\:\!a^2(R-a)\left[\:\!
8\:\!a^2R^4-4\:\!a(4\:\!a^2+1)R^3+(8\:\!a^4-8\:\!a^2-1)R^2+12a^3 R+3\:\!a^2
\:\!\right]}{R^2 \left[\:\!
2\:\!a(R-a)+1
\:\!\right]^4
F},
\end{align}
respectively. 
The sign of $h_0(\rho, z_0)$ on $\gamma_0$ is also determined by that of $G(R)$.

\begin{figure}[t]%[htbp]
\centering
\includegraphics[width=15.7cm,clip]{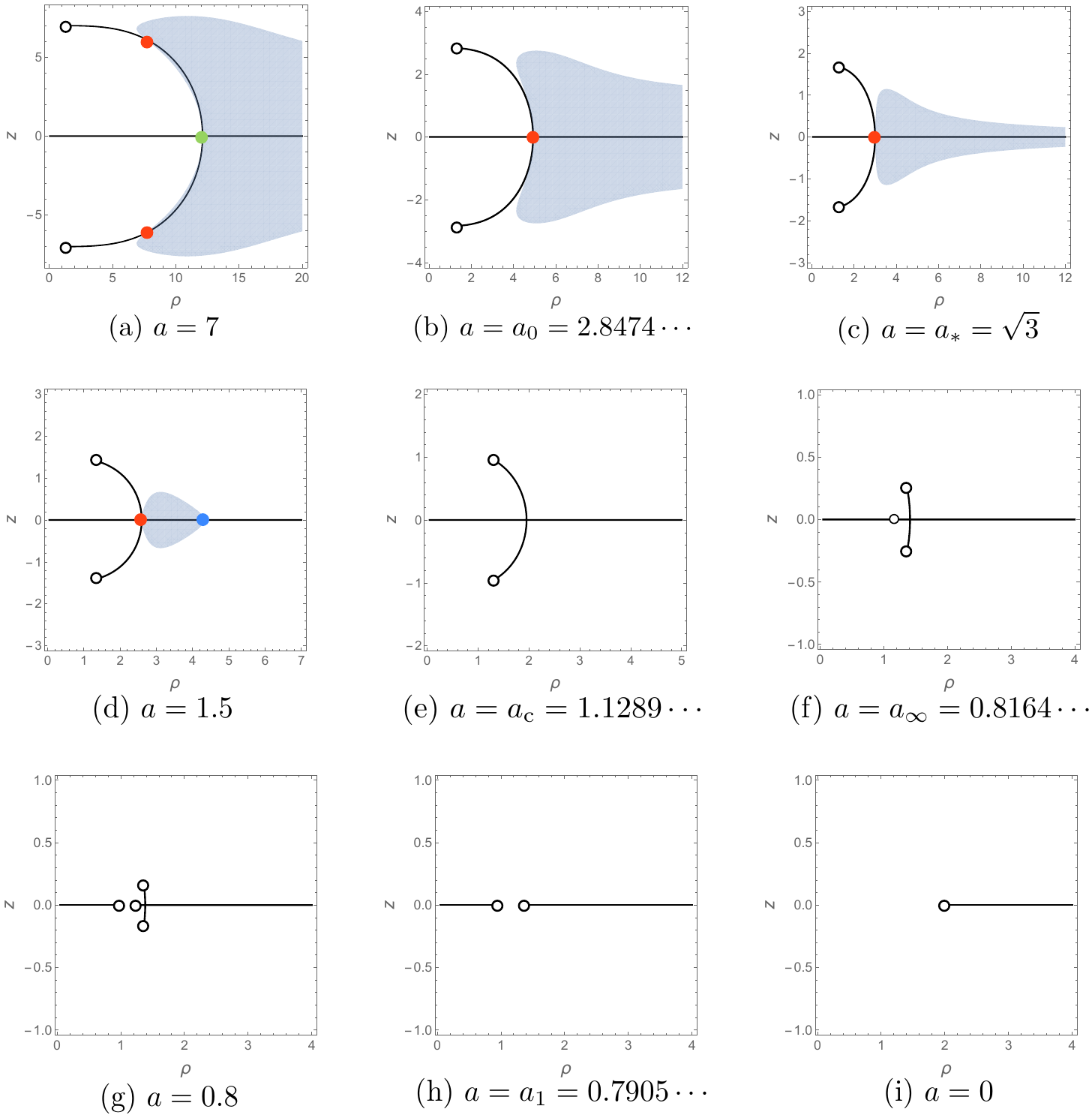}
 \caption{Sequences of circular orbits in the 5D MP dihole spacetimes. 
 Units in which $M=1$ are used. 
Each black solid line shows $\gamma_0$, which is a sequence of circular orbits. 
Each blue shaded region shows $D$, in which circular orbits are stable. 
The boundaries of $D$ are determined by $h_0=0$. 
Points colored by red, green, and blue denote the ISCO, MSCO, and OSCO, respectively. 
Each white point indicates an unstable photon circular orbit. 
(a) $\gamma_0$ between the red and green points overlaps $D$.
(b) $\gamma_0$ between the red and white points does not overlap $D$.}
 \label{fig:d=5}
\end{figure}

For $a\gg1$, 
the sequence of circular orbits is of a typical form 
as shown in Fig.~\ref{fig:d=5}-(a). 
The black solid line shows $\gamma_0$, and the blue shaded region shows the region 
$D$. On $z=0$, 
stable circular orbits exist within the range of $\sqrt{3}\:\!a \leq \rho<\infty$ because 
both $h_0(\rho, 0)$ and $k_0(\rho, 0)$ are positive in this range [i.e., $g(R)\geq 0$]. 
The green point $(\rho, z)=(\sqrt{3}a, 0)$ corresponds to 
a marginally stable circular orbit~(MSCO), 
where $h_0$ vanishes, i.e., $g(R)=0$. 
Furthermore, 
stable circular orbits also appear on the branch $z=z_0$ and 
extend from the MSCO to the ISCOs, which correspond to the red points, 
where $h_0$ also vanishes, i.e., $G(R)=0$. 
On the sequences of stable circular orbits, 
we have $\mathrm{d}E_0/\mathrm{d}R>0$ and $\mathrm{d}L_0/\mathrm{d}R>0$. 
The sequences $\gamma_0$ appearing outside the region $D$ are 
those of unstable circular orbits. 
They are distributed in $0\leq \rho\leq \sqrt{3} a$ on the $z=0$ plane 
and in $R_{\mathrm{L}}<R\leq 2a$ on $z=z_0$, where 
$R_{\mathrm{L}}$ is given as a solution to $F=0$, 
\begin{align}
R_{\mathrm{L}}:=\frac{1}{8\:\!a}\left[\:\!
4\:\!a^2+1+\sqrt{16\:\!a^4+56\:\!a^2+1}
\:\!\right]. 
\end{align}
The positions $(R, z)=(R_{\mathrm{L}}, z_0(R_{\mathrm{L}}))$ 
are shown by white points in Fig.~\ref{fig:d=5}-(a). 
Note that $E_0^2$ and $L_0^2$ become infinite here. 
This corresponds to photon circular orbits 
because the divergence of the unit mass quantities indicates the massless limit 
and the ratio $L_0/E_0$ remains finite even in this limit. 
Consequently, we find that the last circular orbits correspond to 
unstable photon circular orbits. 

As the value of $a$ gradually decreases, 
the ISCOs approach the MSCO. 
Eventually, they merge when $a$ reaches a specific value, $a_0$, as shown in 
Fig.~\ref{fig:d=5}-(b). 
Then the two real roots of $h_0(\rho, z_0)=0$ must be degenerate at $R=2a$, i.e., 
\begin{align}
G(2a)=a (8\:\!a^4-64\:\!a^2-7)=0.
\end{align}
As a solution to this equation, we define
\begin{align}
a_0:=\frac{1}{2}\sqrt{16+3\sqrt{30}}=2.8474\cdots.
\end{align}
When $a$ is within $a_0\geq a \geq a_*$,
where 
\begin{align}
a_*:=\sqrt{3},
\end{align}
the quantities $h_0(\rho, 0)$ and $k_0(\rho, 0)$ are not negative 
in a half-line region $\sqrt{3}\:\!a \leq \rho<\infty$,%
%%%%%
\footnote{If $a\geq \sqrt{3}$, then $g(R)>0$ and $f(R)>0$ in the range $2\:\!a \leq R<\infty$.} 
%%%%%
and hence, stable circular orbits appear there. 
The point $(\rho, z)=(\sqrt{3} a, 0)$ corresponds to the ISCO, 
colored by red in Figs.~\ref{fig:d=5}-(b) and \ref{fig:d=5}-(c), 
where $h_0(\rho, 0)=0$.

Let us now consider the reason why stable circular orbits exist at infinity for $a\geq a_*$. 
The asymptotic expansion of $V(\rho, 0)$ at $\rho\to \infty$ is given by
\begin{align}
\label{eq:expa}
V(\rho, 0)-1=-\frac{4-L^2}{\rho^2}+\frac{
4(a^2-3)+6(4-L^2)}{\rho^4}+O(\rho^{-6}).
\end{align}
If $L^2<4$, 
the leading term, the sum of the Newtonian gravitational potential 
and the centrifugal potential, 
is negative, and if 
$4(a^2-3)+6(4-L^2)>0$, 
the subleading term is positive. 
Furthermore, if $0<4-L^2\ll 1$, then we find 
a local minimum point of $V(\rho, 0)$ in the asymptotic region, 
\begin{align}
\rho\simeq 2\sqrt{\frac{2(a^2-3)}{4-L^2}}.
\end{align}
In order for this value to be a nonzero real number, 
we have $a>a_*$. 
For the marginal case $a=a_*$, we also find a local minimum point of $V(\rho, 0)$ in the asymptotic region,%
\footnote{
The expansion~\eqref{eq:expa} around $a=a_*$ is 
\begin{align}
\label{eq:Veps}
V(\rho, 0)-1
&=
\sum_{l=1}^{\infty}V_{(2l)},\quad
V_{(2)}=-\epsilon/\rho^2, \quad 
V_{(4)}=6\:\!\epsilon/\rho^4, \quad 
V_{(6)}=14(2-3\epsilon)/\rho^6,
\end{align}
where $\epsilon=4-L^2>0$. 
Comparing each term in the limit $\rho\to \infty$ and $\epsilon \to 0$, 
we have
\begin{align}
\left|\:\!V_{(4)}/V_{(2)}\:\!\right|&%=\frac{6}{\rho^2}
=O(\rho^{-2}),
\quad
\left|\:\!V_{(6)}/V_{(2)}\:\!\right|%=\frac{14(2/\epsilon-3)}{\rho^4}
=O(\epsilon^{-1}\rho^{-4}),
\quad
\left|\:\!V_{(6)}/V_{(4)}\:\!\right|%=\frac{7 (2/\epsilon -3)}{3\:\!\rho^2}
=O(\epsilon^{-1}\rho^{-2}).
\end{align}
If $\epsilon$ and $ \rho$ satisfy $\epsilon \rho^4=O(1)$ in this limit, 
$V_{(2)}$ and $V_{(6)}$ are dominant even in the asymptotic region, 
 and $V_{(4)}$ is negligible. 
Then, we find a local minimum point of $V(\rho,0)$ in the asymptotic region, 
$\rho\simeq \sqrt[4]{84}\epsilon^{-1/4}$.} 
and therefore, 
stable circular orbits appear in $3\leq \rho <\infty$ [see Fig.~\ref{fig:d=5}-(c)]. 
This is why we can conclude that there exist stable circular orbits even at infinity 
in the range $a\geq a_*$. 
Note that in the case of a 5D static and spherically symmetric black hole, 
there is no stable circular orbit~\cite{Tangherlini:1963bw, Hackmann:2008tu}. 
This implies that the existence of stable circular orbits is due to the dihole separation, 
and furthermore, in the range of $a\geq a_*$, its effects can be observed at infinity.

When $a<a_*$, there is no longer stable circular orbit at infinity. 
However, 
for $a_*>a>a_{\mathrm{c}}$, 
where $a_{\mathrm{c}}$ is determined by the discussion below, 
a sequence of stable circular orbits draws a segment with finite length on $z=0$, 
as shown in Fig.~\ref{fig:d=5}-(d).
The segment appears in the interval 
\begin{align}
R_{\mathrm{ISCO}}\leq R\leq R_{\mathrm{OSCO}},
\end{align}
where $R_{\mathrm{ISCO}}:=2a$ and $R_{\mathrm{OSCO}}$ correspond to, respectively, 
the radii of the ISCO and the outermost stable circular orbit (OSCO), which 
solve $h_0(\rho, 0)=0$ and 
are denoted, respectively, by the red point and the blue point in Fig.~\ref{fig:d=5}-(d). 
Both radii of the ISCO and the OSCO monotonically decrease as $a$ decreases and 
are degenerate at $a=a_{\mathrm{c}}$, where 
\begin{align}
a_{\mathrm{c}}:=\frac{\sqrt{10+6\sqrt{3}}}{4}=1.1289\cdots,
\end{align}
which are determined by the degenerate condition of 
two roots of $h_0(\rho, 0)=0$ at $R=2a$, i.e., 
\begin{align}
g(2a)=8a^4 (32a^4-40a^2-1)=0. 
\end{align}
The sequences of unstable circular orbits 
are distributed to $0\leq R\leq 2a$ and $R_{\mathrm{OSCO}}<R<\infty$ on $z=0$ and 
to $R_{\mathrm{L}}<R\leq 2a$ on $z=z_0$.

In the range $a\leq a_{\mathrm{c}}$, 
there is no overlapping set of $\gamma_0$ and $D$, i.e., 
there is no stable circular orbit. 
Therefore, we focus only on the $a$-dependent deformation of $\gamma_0$. 
As $a$ decreases from $a_{\mathrm{c}}$ to $a_{\infty}$, 
the outline of $\gamma_0$ remains the same as that in Fig.~\ref{fig:d=5}-(e), 
where 
\begin{align}
a_\infty:=\frac{\sqrt{6}}{3}=0.8164\cdots. 
\end{align}
When $a=a_\infty$, the function $f(R)$ in Eq.~\eqref{eq:f5} 
vanishes only at $R=\sqrt{2}$ (i.e., $\rho=2\sqrt{3}/3$), 
where $E_0$ and $L_0$ diverge. 
This means that there exists an unstable circular orbit not for massive particles 
but for massless particles, which are shown by a white point on the $z=0$ plane in 
Fig.~\ref{fig:d=5}-(f).

In the range $a< a_{\infty}$, 
the sequence of $\gamma_0$ on $z=0$ separates into two parts. 
Note that the outer boundary of the inner sequence and the inner boundary of the outer sequence are unstable photon circular orbits, which are located at 
$R=[\:\!2\pm\sqrt{2(2-3\:\!a^2)}\:\!]^{1/2}$. 
When $a=a_1$, the sequence of $\gamma_0$ on $z=z_0$ vanishes at 
$(R, z)=(2a, 0)$, 
and at the same time the inner boundary of the outer sequence on the $z=0$ plane 
vanishes at the same point, i.e., 
$R_{\mathrm{L}}=2\:\!a=
[\:\!2+\sqrt{2(2-3\:\!a^2)}\:\!]^{1/2}$, 
where 
\begin{align}
a_1:=\frac{\sqrt{10}}{4}=0.7905\cdots.
\end{align}
In $a<a_1$, the sequence on $z=z_0$ no longer appears, and 
the inner and outer sequences on $z=0$ only appear. 
In the limit $a\to0$, the geometry approaches the extremal Reissner-Nordstr\"om 
black hole spacetime with mass $2$. Then 
the inner sequence on $z=0$ disappears, 
and the inner boundary of the outer sequence, the unstable photon circular orbit, 
limits to $(\rho, z)=(2, 0)$ (see the Appendix).

\begin{figure}[t]%[htbp]
\centering
\includegraphics[width=8cm,clip]{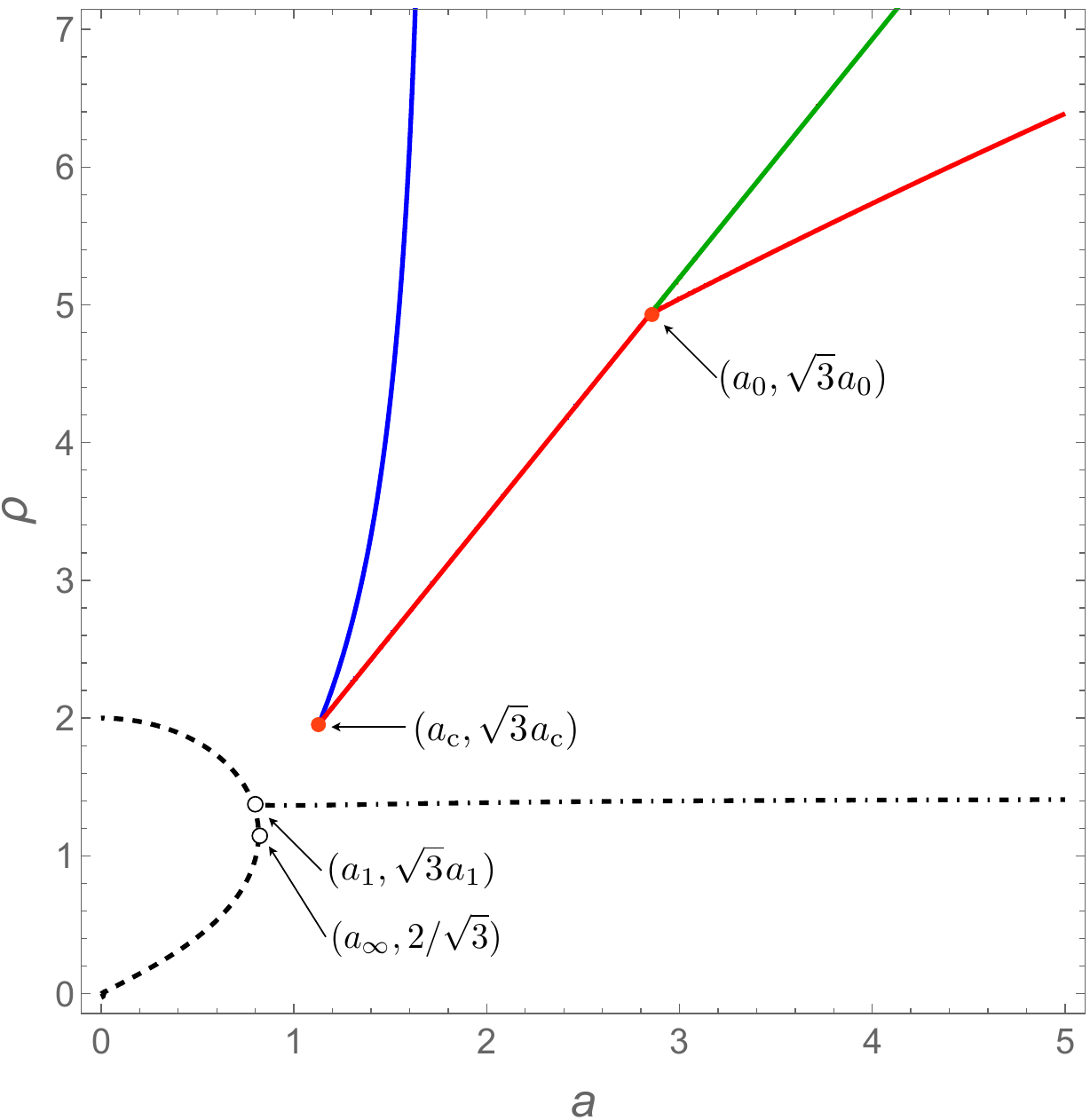}
 \caption{Dependence of characteristic radii on $a$ in the 5D MP dihole spacetime. Units in which $M=1$ are used. 
 The curves colored by green, red, and blue show the MSCO, ISCO, and OSCO, respectively. 
 Note that the OSCO appears only in the range $a_{\mathrm{c}}\leq a<a_*$. 
 Dashed and dot-dashed curves colored by black show 
 unstable photon circular orbits on $z=0$ and $z=z_0$, respectively.}
 \label{fig:CO5}
\end{figure}

We summarize the dependence of characteristic radii on $a$ in Fig.~\ref{fig:CO5}. 

%%%%%
\subsection{$d\geq6$}
%%%%%
We consider the dependence of sequences of circular orbits on the separation 
in the 6D MP dihole spacetime. 
We summarize several quantities associated with circular orbits and their stability. 
From Eqs.~\eqref{eq:E0} and \eqref{eq:L0}, 
$E_0^2$ and $L_0^2$ in $d=6$ are given by
\begin{align}
E_0^2(\rho, 0)&=\frac{R^6(R^5+2\:\!a^2)}{(R^3+2)^2f},
\\
L_0^2(\rho, 0)&=\frac{6(R^2-a^2)^2(R^3+2)^{2/3}}{R^2 f},
\end{align}
respectively, where 
\begin{align}
f(R):=R^5-6R^2+8\:\!a^2.
\end{align}
From Eqs.~\eqref{eq:dE0} and \eqref{eq:dL0}, 
the derivatives of $E_0(\rho, 0)$ and $L_0(\rho, 0)$ take the form 
\begin{align}
\frac{\mathrm{d}E_0(\rho, 0)}{\mathrm{d} R}
&=\frac{3g}{R(R^3+2)(R^5+2\:\!a^2)f},
\\
\frac{\mathrm{d}L_0(\rho, 0)}{\mathrm{d} R}
&=\frac{g}{2R(R^3+2)(R^2-a^2)f},
\end{align}
respectively, where 
\begin{align}
g(R):=-2(R^2-a^2)\left[\:\!
5R^2(R^3+2)+2f
\:\!\right]
-R^2(R^2-5\:\!a^2)(R^3+2)^2.
\end{align}
From Eqs.~\eqref{eq:h0} and \eqref{eq:k0}, the quantities $h_0$ and $k_0$ take the forms 
\begin{align}
h_0(\rho, 0)&=\frac{144 R^6 (R^2-5\:\!a^2)g}{(R^3+2)^6 f^2},
\\
k_0(\rho, 0)&
=-\frac{24 R^2 
(R^2-a^2)\left[\:\!
5R^2(R^3+2)+2f\:\!\right]
}{(R^3+2)^4f},
\end{align}
respectively. 
Besides $z=0$, 
there exists the following branch that satisfies $U_z=0$: 
\begin{align}
z=z_0(R). 
\end{align}
Unlike Eq.~\eqref{eq:z0} in the case $d=5$, however, 
it is not possible to write $z_0$ explicitly in this case because 
the condition is given as an algebraic equation of the fifth degree or more.

\begin{figure}[t]%[htbp]
\centering
\includegraphics[width=11.3cm]{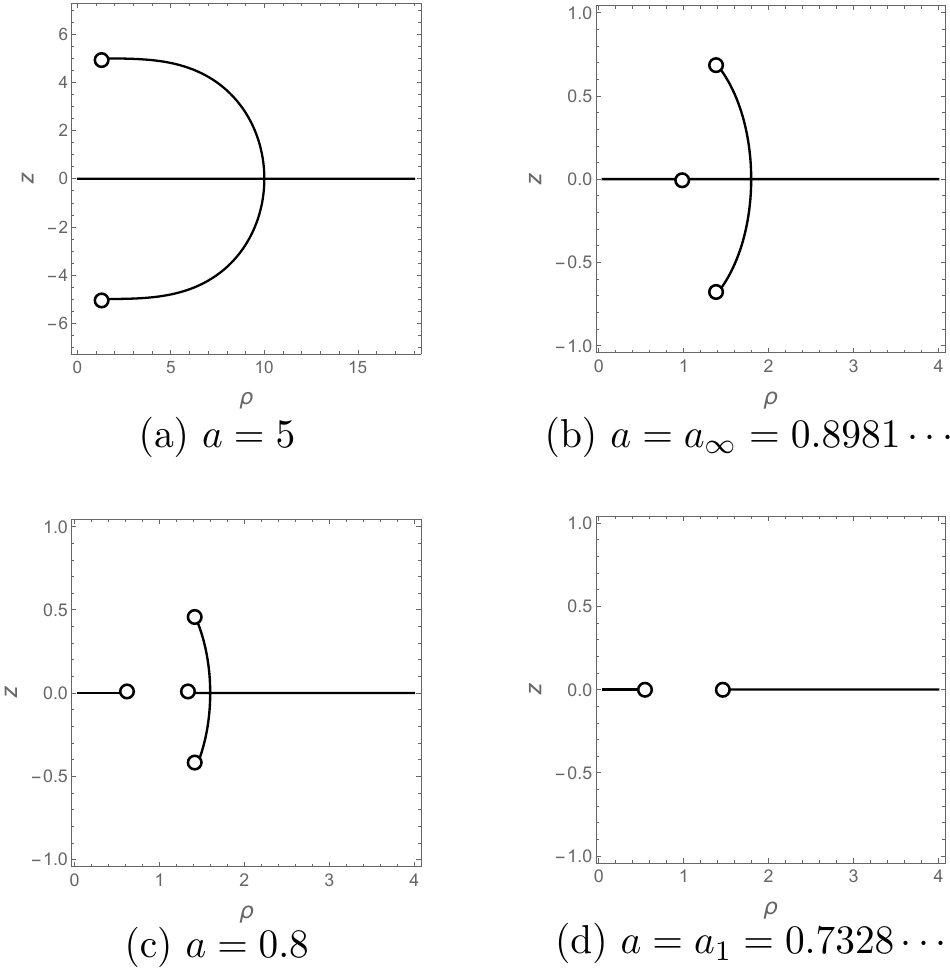}
 \caption{
 Sequences of circular orbits in the 6D MP dihole spacetimes. Units in which $M=1$ are used. 
Each black solid line shows $\gamma_0$. 
Each white point indicates an unstable photon circular orbit. 
}
 \label{fig:d=6}
\end{figure}

In the case $a\gg 1$, the sequences of circular orbits typically show 
the shape depicted in Fig.~\ref{fig:d=6}-(a), where 
black solid curves are $\gamma_0$. 
The set $\gamma_0$ contains $z=0$ and a part of $z=z_0$.
The angular momentum $L_0^2$ and energy $E_0^2$ 
diverge at the boundaries of $\gamma_0$ on $z=z_0$ 
corresponding to the two white points, 
where there exist unstable circular orbits for massless particles 
rather than massive particles. 
As is seen from the asymptotic expansion of $V(\rho, 0)$ at $\rho\to \infty$, 
\begin{align}
V(\rho, 0)-1=\frac{L^2}{\rho^2}-\frac{4}{\rho^3} 
+O(\rho^{-5}),
\end{align}
the leading term ``the centrifugal potential" 
and the subleading term ``the Newtonian gravitational potential" do not make a potential well, so that 
there is no stable circular orbit in the asymptotic region. 
The nonexistence of stable circular orbits is not only in the asymptotic region 
but in the whole region. 
In fact, unlike in $d=4, 5$, the region $D$ does not appear in $d=6$. 
We can interpret that the centrifugal force barrier at the center 
is ineffective to make a local minimum of $V$.

When $a$ takes the value
\begin{align}
a_\infty:=\frac{3}{10}\sqrt[6]{720}=0.8981\cdots,
\end{align}
there appears a (white) point on $z=0$ [corresponding to $R=R_\infty:=
\sqrt[3]{12/5}
$ (i.e., $\rho=\rho_\infty:=\sqrt[6]{720}\sqrt{11}/10$] such that $f(R)$
vanishes, i.e., $E_0^2$ and $L_0^2$ diverge. 
This indicates the 
appearance of an unstable photon circular orbit.

As $a$ is further decreasing, 
the outer boundary of the inner sequence approaches the $z$ axis, 
whereas the inner boundary of the outer sequence goes away from the $z$ axis [see Fig.~\ref{fig:d=6}-(c)], 
and at $a=a_1$, where 
\begin{align}
a_1:=
\dfrac{\sqrt[3]{22}\sqrt[6]{5}}{5}=0.7328\cdots,
\end{align}
it coincides with the point $(\rho, z)=(2a_1, 0)$, where $ \rho_1=2\:\!a_1$. 
At the same time, the boundaries of $\gamma_0$ on $z=z_0$ also limits 
to the same point. 
In other words, 
the three unstable photon circular orbits are degenerate there~[see Fig.~\ref{fig:d=6}-(d)].

For $a\geq a_1$, $\gamma_0$ contains only two sequences on $z=0$. 
In the limit $a\to 0$, 
the inner sequence vanishes at the origin,
and the inner boundary of the outer sequence limits to $(\rho, z)=(2, 0)$ (see the Appendix).

\begin{figure}[t]%[htbp]
\centering
\includegraphics[width=16.0cm]{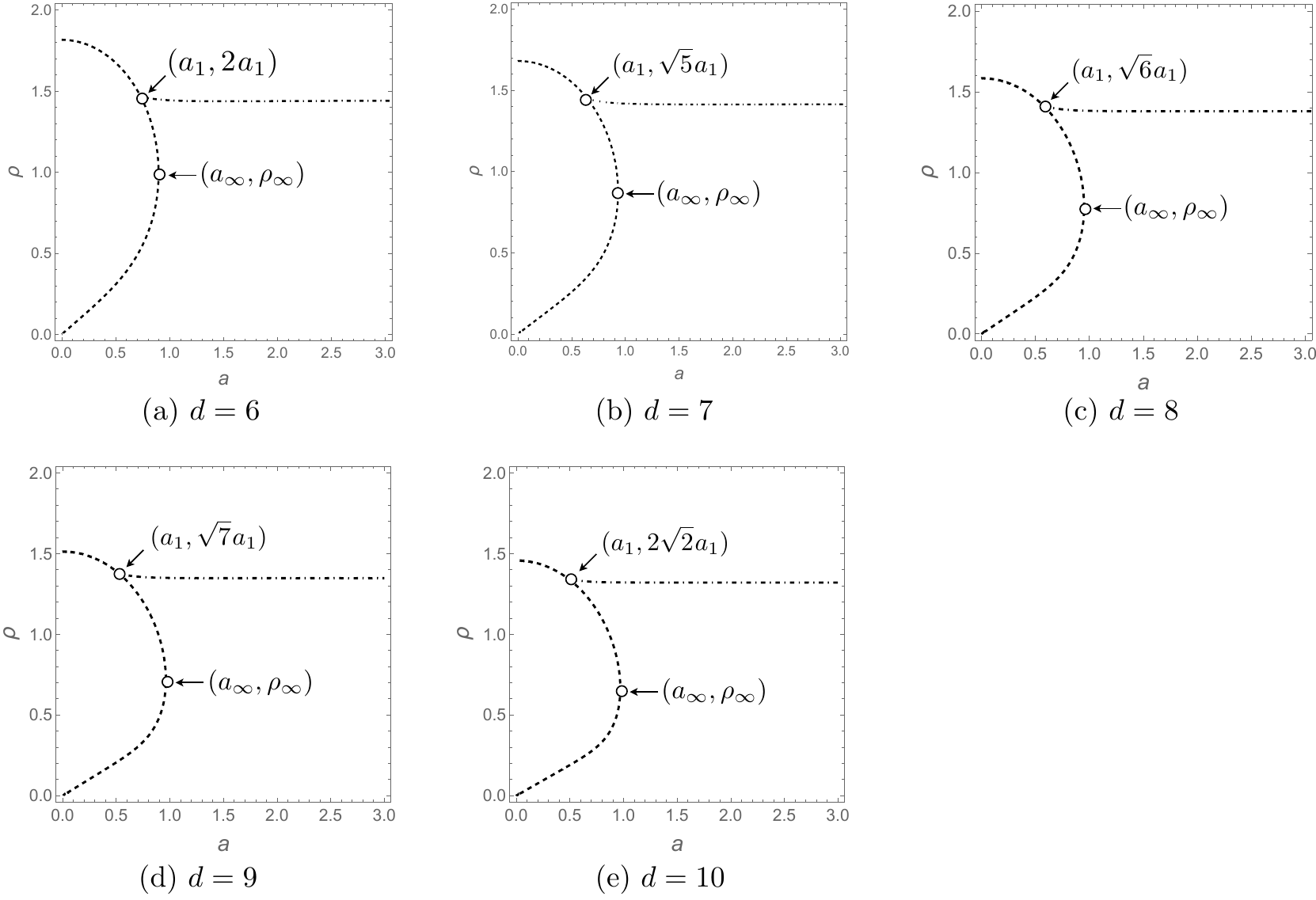}
 \caption{
 Dependence of characteristic radii on $a$ in $d=6, 7, \ldots, 10$. 
 Units in which $M=1$ are used. 
 Dashed and dot-dashed curves colored by black show 
 unstable photon circular orbits on $z=0$ and $z=z_0$, respectively. 
}
 \label{fig:PCO}
\end{figure}

Finally we comment on how the $a$ dependence of 
the sequences of circular orbits changes as $d$ increases. 
At least for $d=7, 8, 9, 10$, we have checked that the change 
in the sequences of circular orbits is qualitatively the same as in the case of $d=6$.
Indeed, it can be seen from Figs.~\ref{fig:PCO} that the dependence of 
the radii of unstable photon circular orbits on $a$ is qualitatively the same for 
$d=6, 7, \ldots, 10$.
The critical values of the separation and the radius, $a_{\infty}$, $ \rho_{\infty}$, $a_1$, and $ \rho_1$, which are defined in the same sense as the $d=6$ case, are summarized in Table~\ref{table:1}. 
Note that $\rho_1=\sqrt{d-2} a_1$. 
This result suggests that the qualitative properties of stable/unstable circular orbits 
are common in the MP dihole spacetime in $d\geq 6$. 
\begin{table}[t]
\begin{tabular}{lllll}
\hline\hline
$d$~~~~~~~~~&$a_1$&$\rho_1$&$a_{\infty}$&$\rho_{\infty}$
\\
\hline
\\[-5mm]
$6$&
$\dfrac{\sqrt[3]{22}\sqrt[6]{5}}{5}=0.7328\cdots$&
$2\:\!a_1=1.4656\cdots$&$\dfrac{3}{10}\sqrt[6]{720}=0.8981\cdots$&
$\dfrac{\sqrt[6]{720} \sqrt{11}}{10}=0.9929\cdots$
\\[2mm]
$7$&
$\dfrac{\sqrt{2}\sqrt[4]{57}}{6}=0.6476\cdots$~~~~~&
$\sqrt{5}\:\!a_1=1.4481\cdots$&
$\dfrac{4\sqrt[4]{150}}{15}=0.9332\cdots$&
$\dfrac{\sqrt{35}\sqrt[4]{24}}{15}=0.8729\cdots$
\\[2mm]
$8$&
$\dfrac{\sqrt[5]{58}}{\sqrt[10]{7^7}}=0.5769\cdots$&
$\sqrt{6}\:\!a_1=1.4131\cdots$&
$\dfrac{5\sqrt{3}\sqrt[5]{5}\sqrt[10]{7^3}}{21\sqrt[10]{2}}=0.9517\cdots$&
$\dfrac{\sqrt[5]{5}\sqrt[10]{7^3}\sqrt{51}}{\sqrt[10]{2} 21}=0.7848\cdots$
\\[2mm]
$9$&
$\dfrac{\sqrt[6]{82}}{4}=0.5210\cdots$&
$\sqrt{7}\:\!a_1=1.3786\cdots$&
$\dfrac{3\sqrt[6]{3}}{\sqrt{14}}=0.9628\cdots$&
$\sqrt{\dfrac{5}{14}}3^{1/6}=0.7176\cdots$
\\[2mm]
$10$&
$\dfrac{1}{3}\sqrt[7]{\dfrac{110}{9}}=0.4766\cdots$&
$2\sqrt{2}\:\!a_1=1.3481\cdots$&
$\dfrac{7\sqrt[7]{7}}{6\sqrt[14]{8}\sqrt[7]{9}}=0.9701\cdots$&
$\dfrac{\sqrt[7]{7} \sqrt{23}}{6\sqrt[14]{8}\sqrt[7]{9}}=0.6646\cdots$
\\[2mm]
\hline\hline
\end{tabular}
\caption{Critical values of $a_1$, $\rho_1$, $a_{\infty}$, and $\rho_\infty$ for $d=6, 7, 8, 9, 10$. }
\label{table:1}
\end{table}

%%%%%%%%%%
\section{Summary and discussions}
\label{sec:4}
%%%%%%%%%%

We have considered the dynamics of particles, 
focusing on circular orbits, in the $d$-dimensional MP dihole spacetime~($d\geq 5$).
Using the on-shell conditions for geodesic motion, 
we have clarified the conditions for the existence of circular orbits 
in terms of a 2D effective potential 
and have also provided a prescription for determining whether these orbits are stable.
Applying this formalism to the case of $d=5$, 
we have shown the dependence of sequences of stable/unstable circular orbits 
on the dihole separation. 
One of the most remarkable features is the appearance of stable circular orbits
because it was shown in the previous works 
that they were not found in single black holes with 
a spherical horizon~\cite{Tangherlini:1963bw, Hackmann:2008tu}. 
Particularly for the large separation $a\geq a_*=\sqrt{3}$, 
they appear from the ISCO to infinity, whereas for 
$a_{\mathrm{c}}(=1.1289\cdots)<a<a_*$, they 
exist only in a restricted region between the ISCO and the OSCO. 
Therefore, we can interpret this phenomenon 
as a result caused by the existence of two horizons. 
In other words, 
the center of the system is shifted off the horizon, 
and as a result, the effect of the centrifugal barrier near the center affects 
the existence of stable circular orbits in the intermediate region.
Furthermore, the existence of the stable circular orbits in the asymptotic region is 
also affected by the power law of gravitational force specific to 5D.
Let us note that as shown in Ref.~\cite{Wunsch:2013st, Nakashi:2019mvs} 
for the 4D MP dihole spacetime, stable circular orbits exist in the asymptotic region 
for arbitrary separations. 
Therefore, we can conclude that the appearance of the OSCO is a proper phenomenon of 5D.

In the cases of $d=6, 7, \ldots, 10$, we have found that there is no 
stable circular orbit for any values of $a$. 
These results suggest that stable circular orbits do not appear for $d\geq 6$ in general. 
On the other hand, how the sequences of circular orbits change 
by the separation $a$ for $d\geq 6$ is qualitatively the same as $d=5$. 
We expect that this property is also dimension independent in $d\geq 6$.
It is worth noting that though the existence of a stable photon circular orbit is 
one of the specific geodesic structures in the 4D MP dihole spacetime, 
it is absent in $d\geq 5$.

As seen in the Schwarzschild-Tangherlini and Myers-Perry black hole backgrounds, 
stable circular orbits tend not to appear for any dimensions of $d\geq 5$. 
This seems to be a property common to black objects with a large value of $d$.
In fact, even in the MP dihole spacetime, we have found that stable circular orbits tend to be absent for any dimensions of $d\geq 6$. 
However, our results in 5D imply that there is a mechanism for the existence of stable circular orbits even in higher dimensions due to many-body effects, i.e., the existence of the center outside the horizon.

In contrast to the fact that the metric is analytic 
on the horizon of the 4D MP multi-black hole~\cite{Hartle:1972ya}, 
the horizons of higher-dimensional ones are generally not 
smooth~\cite{Welch:1995dh, Candlish:2007fh}. 
For $d=5$, 
the metric can be $C^2$ on the horizon but cannot be $C^3$ in general.
For $d>5$, the metric is not even $C^2$ on the horizon, 
which leads to unavoidable curvature singularities. 
Hence, it should be noted that the results we have obtained in the case of $d\geq 6$ 
may include the effect of singularities.

For comparison with observations, 
we should discuss a higher-dimensional model of the universe, 
e.g., a higher-dimensional black hole spacetime in which 
the extra dimensions are compactified.
In particular, a 5D Kaluza-Klein black hole spacetime with a twisted $S^1$ fiber behaves as 
a 5D spacetime near the horizon ($S^3$ topology), 
whereas it effectively behaves as a 4D flat spacetime 
(i.e., a 4D flat spacetime with a compact dimension) 
in the asymptotic region~\cite{Tomizawa:2011mc, Tomizawa:2018syg}. 
From this point of view, the 5D Kaluza-Klein black holes connect 4D spacetimes 
and 5D spacetimes as well as they have the feature of both 4D and 5D. 
Therefore, we expect that 
the size of the extra dimension should affect the sequence of stable circular orbits. 
This is an interesting issue for the future.

\begin{acknowledgments}
This work was supported by the Grant-in-Aid for Early-Career Scientists~[JSPS KAKENHI Grant No.~JP19K14715 (T.I.)] and 
Grant-in-Aid for Scientific Research (C) [JSPS KAKENHI Grant No.~JP17K05452 (S.T.)]
from the Japan Society for the Promotion of Science. 
S.T. is also supported from Toyota Institute of Technology Fund for
Research Promotion A.
\end{acknowledgments}

\appendix
%%%%%%%%%%
\section{Unstable photon circular orbit in the extremal Reissner-Nordstr\"om spacetime}
\label{sec:A}
%%%%%%%%%%
We review an unstable photon circular orbit in the $d$-dimensional extremal Reissner-Nordstr\"om spacetime. 
The metric is given in isotropic coordinates by
\begin{align}
\label{eq:RNiso}
\mathrm{ds}^2
=-\left(
1+\frac{M}{r_*^{d-3}}
\right)^{-2}\:\!\mathrm{d}t^2
+\left(
1+\frac{M}{r_*^{d-3}}
\right)^{2/(d-3)} \left(
\mathrm{d}r_*^2+r_*^2 \:\!\mathrm{d}\Omega^2_{d-2}
\right),
\end{align}
where $M$ is a mass parameter of the extremal black hole. This metric is derived 
by choosing 
$M_+=M_-=M/2$, $a=0$, and $r_*=\sqrt{z^2+\rho^2}$ in Eqs.~\eqref{eq:met}--\eqref{eq:rpm}. 
In the Schwarzschild radial coordinate, 
\begin{align}
r^{d-3}=r_*^{d-3}+M,
\end{align}
the metric~\eqref{eq:RNiso} is written in the standard form 
\begin{align}
\mathrm{ds}^2=-
\left(
1-\frac{M}{r^{d-3}}
\right)^{2}
\:\!\mathrm{d}t^2
+
\left(
1-\frac{M}{r^{d-3}}
\right)^{-2}
\:\!
\mathrm{d}r^2
+r^2\:\!\left[\:\!
\mathrm{d}\theta^2+\sin^2\theta \:\!
\left(\mathrm{d}\phi^2+\sin^2\phi \:\!\mathrm{d}\Omega^2_{d-4}\right)
\:\!\right].
\end{align}
We consider the radial motion of a massless particle. Without loss of generality, 
we assume that a particle is confined to the equatorial plane, $\theta=\pi/2$, 
and has an unique nonzero component of angular velocities $\dot{\phi}\neq0$. 
Let us introduce $E=-(1-M/r^{d-3})^2\dot{t}$, a conserved particle energy, and $L=r^2 \dot{\phi}$, a conserved angular momentum. 
From the on-shell condition for null geodesics, $g_{\mu\nu}\dot{x}^\mu \dot{x}^\nu=0$, 
the radial equation in terms of these constants 
is given by 
\begin{align}
&\dot{r}^2+V=(E/L)^2,
\\
&V=\left(
\frac{1}{r}-\frac{M}{r^{d-2}}
\right)^2,
\end{align}
where we have rescaled the affine parameter. The derivative of the potential function $V$ 
takes the form
\begin{align}
V'=-\frac{2}{r^{2d-3}} \left(r^{d-3}-M\right)\left[\:\!
r^{d-3}-(d-2) M
\:\!\right]. 
\end{align}
From $V'=0$, we find that an extremum point of $V$ is located at 
\begin{align}
\label{eq:eRNph}
r=\sqrt[d-3]{(d-2)M}. 
\end{align}
Here exists an unstable circular orbit
because $V$ has a local maximum point.
In terms of $\rho$, the radius~\eqref{eq:eRNph} becomes
\begin{align}
\rho
=\sqrt[d-3]{(d-3) M}.
\end{align}

\end{document}